# Sub-ppb aerosol detection at a distance of 30 meters by millijoule femtosecond laser pulse filamentation in air

Jiewei Guo [a,b,c,#], Zhi Zhang [a,b,c,#], Nan Zhang [a,b,c*], Binpeng Shang [a,d], Jiayun Xue [a,d], Yuezheng Wang [a,c], Shishi Tao [a,c], Bofu Xie [a,c], Lanjun Guo [a,b,d], Lie Lin [a,b,d], Weiwei Liu [a,b,c*]

[a] Institute of Modern Optics, Eye Institute, Nankai University, Tianjin 300350, China

[b] Tianjin Eye Hospital, Tianjin 300020, China

[c] Tianjin Key Laboratory of Micro-scale Optical Information Science and Technology, Tianjin 300350, China

[d] Tianjin Key Laboratory of Optoelectronic Sensor and Sensing Network Technology, Tianjin 300350, China

[#] Equal contributors

[*] Corresponding author. E-mail: zhangn@nankai.edu.cn, liuweiwei@nankai.edu.cn

**ABSTRACT** In this work, sub-ppb aerosol detection is achieved by femtosecond laser filament with a single pulse energy of 4 mJ at a distance of 30 m. A concave mirror with an open aperture of 41.4 cm is employed in an off-axis optical system to focus the femtosecond laser beam and collect the fluorescence of NaCl aerosol. The simulation and experimental results show that the astigmatism can be greatly reduced when femtosecond laser beam is incident non-symmetrically on the concave mirror. Compared with the case that femtosecond laser strikes at the center of the concave mirror, the intensity of the optical filament is increased by 69.5 times, and the detection of limit of sodium chloride aerosol is reduced by 86%, which is down to 0.32 ppb. The improved excitation scheme in this work utilizes the nonsymmetrical beam spot on the concave mirror to compensate the non-symmetry induced by the off-axis setup, reducing the astigmatism of the focusing laser beam and improving the aerosol's detection of limit.

## 1. Introduction

During the propagation of high intensity femtosecond laser pulses in transparent media, the laser beam can overcome the natural diffraction and form a plasma channel with a diameter of ~100 $um$, which is termed of optical filament[1-4]. The formation of the laser filament can be attributed to the dynamic balance among the beam diffraction, the optical Kerr effect induced self-focusing and the defocusing by the plasma[5]. Femtosecond laser filament has a nearly constant laser intensity of about $10^{13}$ ~ $10^{14}$ W/cm$^2$ [6, 7], which is sufficient to cause the ionization and fragmentation of molecules[8]. The ionization or dissociation of molecules or atoms emits the fingerprint fluorescence spectrum during the relaxation process, which provides the capability of detecting the chemical composition in a long distance[9-12].

Compared with the current methods of atmospheric aerosol composition detection, such as ion chromatography (IC)[13], gas chromatography (GC)[14], atomic absorption spectrometry (AAS)[15], mass spectrometry (MS)[16], etc., femtosecond laser filament-induced plasma spectroscopy (FIPS) can realize real-time remote sensing of the chemical composition of air pollutants in different forms, such as solid, aerosol, gas, etc., which has aroused widespread research interest. In previous works, Daigle et al.[17] reported that the detection limit was ~33ppm with femtosecond laser pulses of 72 mJ at a distance of 50 m. Then, Daigle et al.[18] reported the detection limits for different constituents in aerosol: 127 mg/L (127 ppm) for Fe, 27 mg/L (27 ppm) for Cu, 9 mg/L (9 ppm) for Pb, and 3mg/L (3 ppm) for Na. Recently, using a femtosecond laser with relatively low pulse energy (4.4 mJ), Golik et al.[19] measured the filament-induced fluorescence of aerosols containing Al, Ba, Na, etc. The detection limit of Na was 0.7 mg/L (0.7 ppm) at a distance of 0.5 m. Figure 1 compares the detection limit of Na measured in this work and the detection limits achieved in literatures. It should be noted that the limit of detection reported in literatures is presented in the form of the mass ratio of the metal element in the water droplet (solution concentration). For the convenience of comparison, the limit of detection obtained in this work is also presented by the solution concentration (0.025 ppm, corresponding to 0.32 ppb in air) in Figure 1.

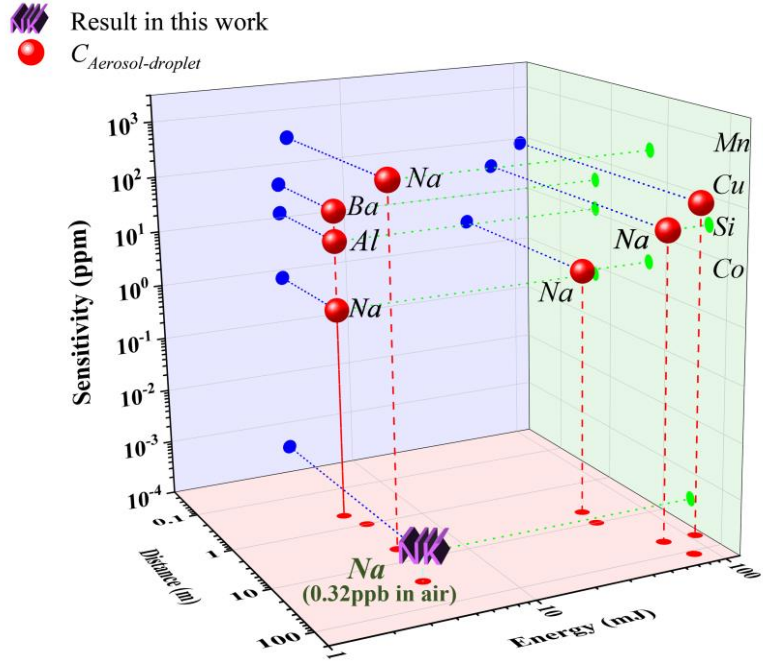

**Fig. 1.** Reported detection limits with different laser pulse energies at different distances.[17-23] The red ball represents the measurement results reported in the literatures, and the "NK" logo represents the measurement result in this work.

The femtosecond laser is considered an attractive LIDAR technology for real-time detection of atmospheric aerosol composition due to its unique filamentation properties in the atmosphere. However, improving the detection sensitivity is still a technical problem faced by this technology. The self-focusing distance is proportional to the square of the beam diameter[17], but in fact, femtosecond laser filamentation system composed of refraction optical components is difficult to achieve large diameter, light weight and low cost[24, 25]. Although reflective femtosecond laser filamentation system has the advantages of large aperture, no chromatic aberration, light weight and low cost which is widely used in atmospheric remote sensing and Earth observation[26-28], astigmatism is inevitably introduced for the off-axis optical system composed of a concave lens and a large diameter concave mirror. Therefore, a simple and efficient scheme that can reduce the astigmatism of the off-axis system is badly needed for increasing the intensity of the filament. Daigle et al. replaced the concave lens with a deformable mirror to correct the wavefront aberration in a closed-loop system[29]. However, the low energy efficiency and low damage threshold of deformable mirrors limit the application in the remote sensing by high intensity femtosecond laser filamentation. Spatial light modulators (SLM) are also used to correct the astigmatism[30-32]. However, SLM's low laser damage threshold limits its operation in high-power laser systems. Recently, Tao et al. proposed to eliminate astigmatism by designing a phase plate with free-form surface[33], whereas one phase plate is only suitable for certain optical setup with fixed focal length.

In this paper, we found experimentally that the astigmatism in the off-axis femtosecond laser filamentation system can be reduced by breaking the symmetry of the beam spot distribution on the concave mirror. The numerical simulations demonstrate that most of the wavefront distortion due to the off-axis configuration can be corrected by the non-symmetrical incidence of the laser beam on the concave mirror. After compensating the astigmatism, a filament was generated at a distance of 30 m, which is practically limited by the lab size. The intensity of the optical filament is increased by 69.5 times, and the detection of limit of sodium chloride aerosol is reduced by 86%, which is down to a record of 0.32 ppb in air, corresponding to 0.025 ppm (mass concentration) of $Na^+$ in aerosol droplets.

## 3. 2. Setup design and numerical simulations

To enhance the filament intensity and improve the detection limit of aerosol, the astigmatism of the off-axis femtosecond laser filamentation system must be reduced as much as possible. The off-axis optical setup is simulated using Zemax software. In Figure 2a the collimated laser beam is focused at a distance of 30 m using the lens group composed of a plano-concave lens ($\Phi$ = 25 mm, $f$ = -150 mm) and a concave mirror ($\Phi$ = 41.4 cm, $f$ = 2 m). The divergent laser beam after passing through the concave lens strikes on the concave mirror at an incident angle of 2.5°. Further deceasing the incident angle will cause the laser beam focused by the concave mirror be blocked by the beam steering mirror (M1 in Figure 2a). The beam spot diameter $(1/e^2)$ on the concave mirror is 20.2 cm. When the beam spot on the concave mirror is symmetric relative to the center of the concave mirror, obvious astigmatism appears near the focal spot as is shown in the insets of Figure 2a. Sagittal beam and tangential beam converge before and after the geometric focus of the system, respectively. To quantitatively present the astigmatism of the setup in Figure 2a, the dependences of the beam diameters $(1/e^2)$ along $X$ and $Y$ directions on the laser propagation distance were calculated and shown in Figure 2b. From Figure 2b, it is found that the distance between the sagittal and tangential focus lines is 40 cm for the setup in Figure 2a. With the help of Zemax software, the wavefront phase on the concave mirror can be calculated, which is shown in Figure 2c. It indicates that the non-symmetric wavefront phase exists on the concave mirror leads to the large astigmatism when the beam strikes at the center of the concave mirror.

In order to evaluate the astigmatism of Figure 2a quantitatively, the aberration characteristics of the system were analyzed through the wave aberration of the system, and the zernike fringe polynomial was used to characterize the wave aberration of the system, where the fifth and sixth terms ($Z_5$ and $Z_6$) of the Zernike Fringe polynomials respectively representing the astigmatism in $X$ and $Y$ directions are calculated. $Z5$ is $C_5 \times (P^2 \times \cos(2A))$ and $Z6$ is $C_6 \times (P^2 \times \sin(2A))$, in which $A$ is the angle measured counterclockwise from the local +x axis, $P$ is the normalized radial coordinate, $C_5$ and $C_6$ are astigmatism coefficients. Since the laser beam is incident on the concave mirror obliquely in the $XOZ$ plane, only $C_5$ in the x direction is non-zero, which is calculated to be $1.26\lambda$, and $C_6 = 0$.

It is found numerically that when the concave mirror moves towards -x direction, $C_5$ gradually decreases to zero. When the beam spot is just tangent to the edge of the concave mirror as is shown in Figure 3a, i.e. the concave mirror shifts 10.6 cm towards -x direction, most of the astigmatism can be reduced (see Figure 3b) and the wavefront of the beam spot on the concave mirror is nearly symmetric (see Figure 3c). In this case, $C_5$ and $C_6$ are respectively 0.045λ and 0. The beam profiles at different propagation distances are shown in the insets of Figure 3a. Clearly, the beam quality near the focal spot has been greatly improved.

The main difference between the two optical setups in Figures 2a and 3a is the different relative position between the laser beam spot and the concave mirror which is shown in Figure 4a. Figure 4b illustrates the variations of the astigmatic parameter $β$ along the propagation direction for the optical system with central/edge incidence. The astigmatic parameter $β$ is defined as the ratio of the beam width $W_j (j = X, Y)$ in the $X$ and $Y$ directions. $β = 1$ represents the beam spot is circular. It is seen from Figure 4b that the edge incidence fully optimizes the astigmatism of the off-axis system, and the beam spot has a circular shape.

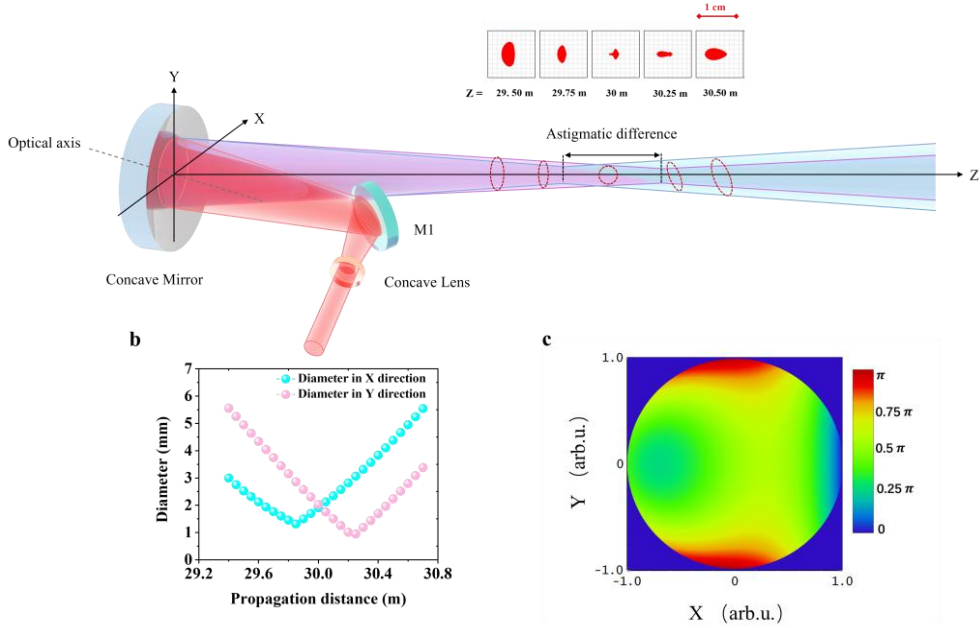

**Fig. 2.** Off-axis reflective filamentation system. The laser beam strikes at the center of the concave mirror. (a) Off-axis reflective setup; (b) dependences of the beam diameters in X and Y directions on the propagation distance; (c) wavefront phase distribution of the incident laser on the concave mirror.

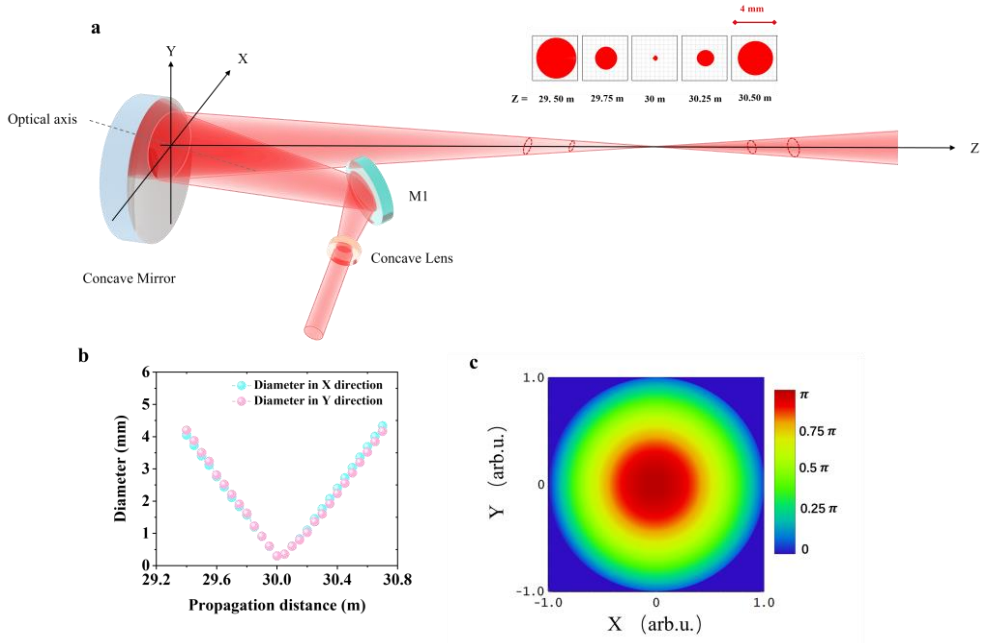

**Fig. 3.** Off-axis reflective filamentation system. The laser beam strikes non-symmetrically on the concave mirror. (a) Off-axis reflective setup; (b) dependences of the beam diameters in *X* and *Y* directions on the propagation distance; (c) wavefront phase distribution of the incident laser on the concave mirror.

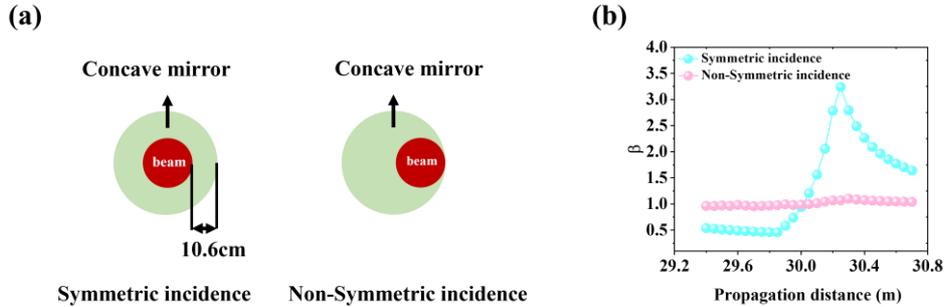

**Fig. 4.** (a) Different relative position between the beam spot and the concave mirror in Figures 2a and 3a; (b) variations of the astigmatic parameter β along the propagation direction.

## 3. Experimental results and discussions

### 3.1. Off-axis reflection system for femtosecond laser filamentation and aerosol detection

In this work, a commercial Ti:Sapphire femtosecond laser system (Legend Elite, Coherent Inc.) was employed to generate 500 Hz, 35 fs, 800 nm, 4 mJ laser pulses. The schematic diagram of the experimental setup is shown in Figure 5a. The laser pulse output from the laser system was focused by a lens group consisting of a concave lens (L1, $\Phi = 25$ mm, $f = -150$ mm) and a concave mirror (L2, $\Phi = 41.4$ cm, $f = 2$ m) which is identical to those used in the numerical simulations. The geometrical focus of the lens group locates 30 m away from the concave mirror (L2). The laser pulse is incident non-symmetrically on one side of the concave mirror after passing through the concave lens, and the reflected laser beam by the concave mirror is focused and form optical filament at a distance of 30 m relative to the concave mirror (L2).

In order to characterize the length and intensity distribution of the optical filament, a microphone (V306, Olympus. Ltd) combined with an amplifier (5072PR, Olympus. Ltd) an oscilloscope (DPO3034, Tektronix Inc.) is used to measure the ultrasonic wave emitted from the optical filament. The typical time domain ultrasonic signal is shown in Figure 5b. Because the length of laser filament (~40 cm) is much longer than the spatial resolution (~0.87 cm) of the ultrasonic microphone, the microphone is mounted on an electrically driven sliding rail and moved parallel to the laser propagation direction to measure the length and distribution of the filament point by point. The spatial step of the microphone is 2.5 cm which is just the spatial resolution of the microphone.

To analyze the residual astigmatism, a CCD camera is used to record the variation of the cross-sectional intensity distribution of the laser beam spot along the beam propagation direction. It should be noted that in order to protect the CCD camera from being damaged, all the beam spots are captured under the linear optical propagation with attenuated pulse energy.

An aerosol generator (HRH-WAG3, Beijing Huironghe Technology Co., Ltd.) is employed to generate sodium chloride aerosol with different concentrations. The aerosol is stably injected into the tube by controlling the air pump of the generator to interact with the filament, which is shown in Figure 5c. The mean diameter of the particle size inside the tube is about 2 $um$ measured by an aerodynamic particle size spectrometer (TSI3321, TSI Inc). The laser filament ionizes the aerosol and generate fingerprint fluorescence which is collected by the concave mirror and focused onto the end of the tail fiber. The backward fluorescence was detected by a grating spectrometer (Omni-λ 300, Zolix Ltd.) equipped with an intensified CMOS camera (Istar-sCMOS, Andor TechnologyLtd.).

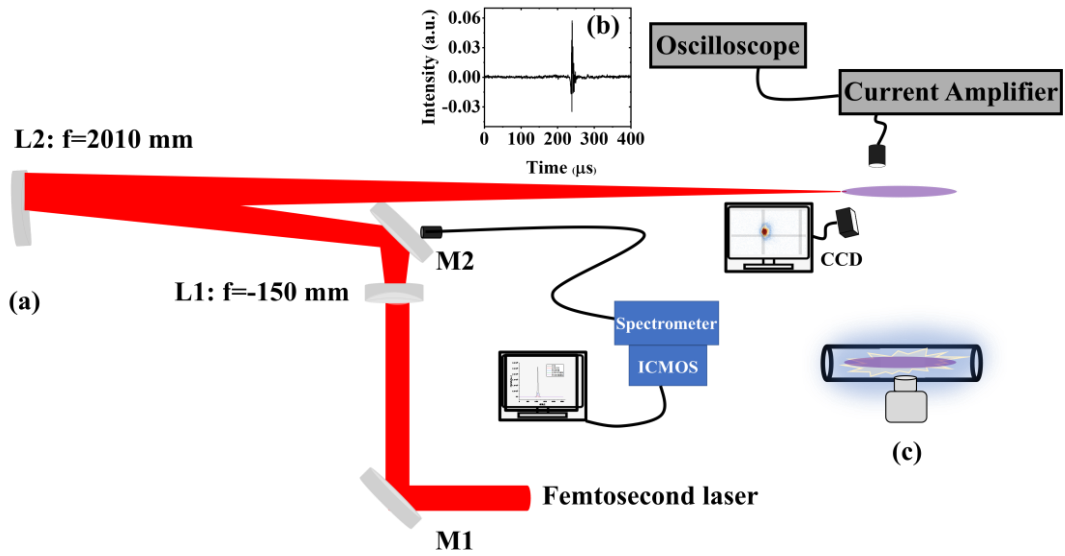

**Fig.5.** (a) Schematic diagram of the experimental setup; (b) typical ultrasonic signal in time domain; (c) aerosol generating device.

*3.2. Improving the limit of detection of aerosol using the astigmatism-compensated off-axis system*

Figures 6a and 6b show the beam profile's variation along the laser propagation direction respectively for the cases that the laser spot strikes at the center or one side of the concave mirror. The diameter $(1/e^2)$ of the laser spot at different positions along the laser propagation direction was extracted and shown in Figures 6c and 6d, which are highly consistent with the simulation results. Experimental results show that the astigmatism is greatly compensated and the distance between the sagittal and tangential focus lines is reduced to be zero when the laser beam non-symmetrically strikes on the concave mirror. Figure 6e illustrates the variations of the astigmatic parameter *β* along the propagation direction, which also agrees well with the simulation results. The astigmatism compensation method proposed here is applicable for different focal lengths, which is superior to the free-form surfaces that only works with one specific focal length [22].

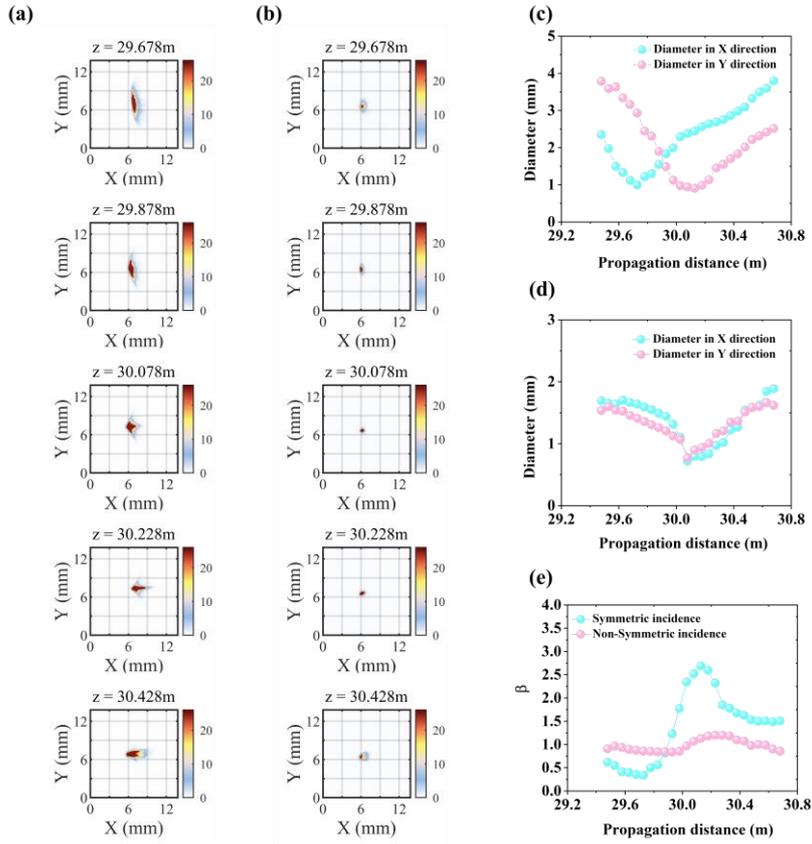

**Fig.6** Variation of beam profiles near the geometric focus along the laser propagation direction recorded by CCD camera. (a) and (c): beam profiles and beam diameters for the case of laser beam incident on the center of the concave mirror; (b) and (d): beam profiles and beam diameters for the case of laser beam incident on one side of the concave mirror; (e) variations of the astigmatic parameter β along the propagation direction.

Figure 7 presents the ultrasonic signal intensity as a function of the laser propagation distance, and the black dotted line represents 3 times standard deviation of the background noises. Due to the astigmatism during the laser filamentation when the laser beam strikes at the center of the concave mirror, the ultrasonic signal of the laser filament has two peaks along the laser filament, respectively representing the sagittal and tangential focal spot. It should be noted that the Y-axis ruler of the pink data is on the right to facilitate analysis. After compensating the astigmatism by making the laser beam non-symmetrically incident on the concave mirror, only one peak is left and its intensity increases by ~70 times, which will greatly enhance the fingerprint fluorescence spectrum of aerosol excited by the laser filament.

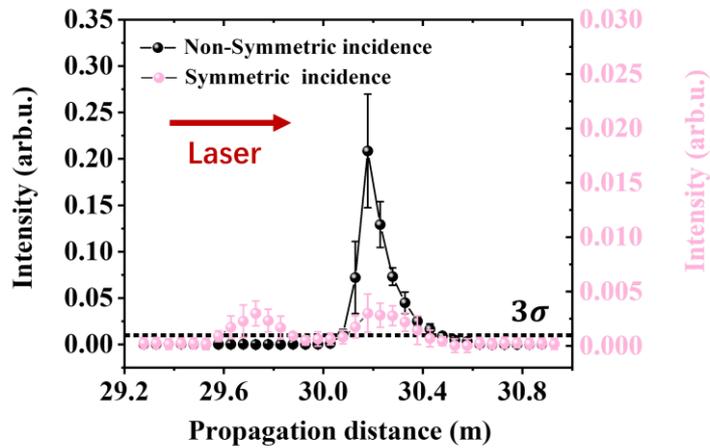

Fig. 7 Ultrasonic signal intensity along the femtosecond laser filament

Figures 8a and 8c show the filament-induced fluorescence spectra of NaCl aerosol with different concentrations recorded respectively by the optical setups in Figures 2a and 3a. The exposure time of 30 s was adopted for all the measurements, and the gain of ICMOS camera is also identical for these spectral curves. Figures 8b and 8d present the integral intensity of fingerprint fluorescence peak as a function of $Na^+$ concentration, which are respectively obtained using the data in Figures 8a and 8c. It can be found that in the logarithmic coordinate system, the relation between the integral intensity and $Na^+$ concentration can be well fitted by a linear function. The intersection point of the fitted line and 3σ line is the detection limit of NaCl aerosol for the setup used in this work. It can be found that in the off-axis reflection femtosecond laser filamentation system, the detection limit can be reduced by 86% when the laser beam spot strikes non-symmetrically on the concave mirror, leading to a sub-ppb limit of detection, which is the lowest one as we know.

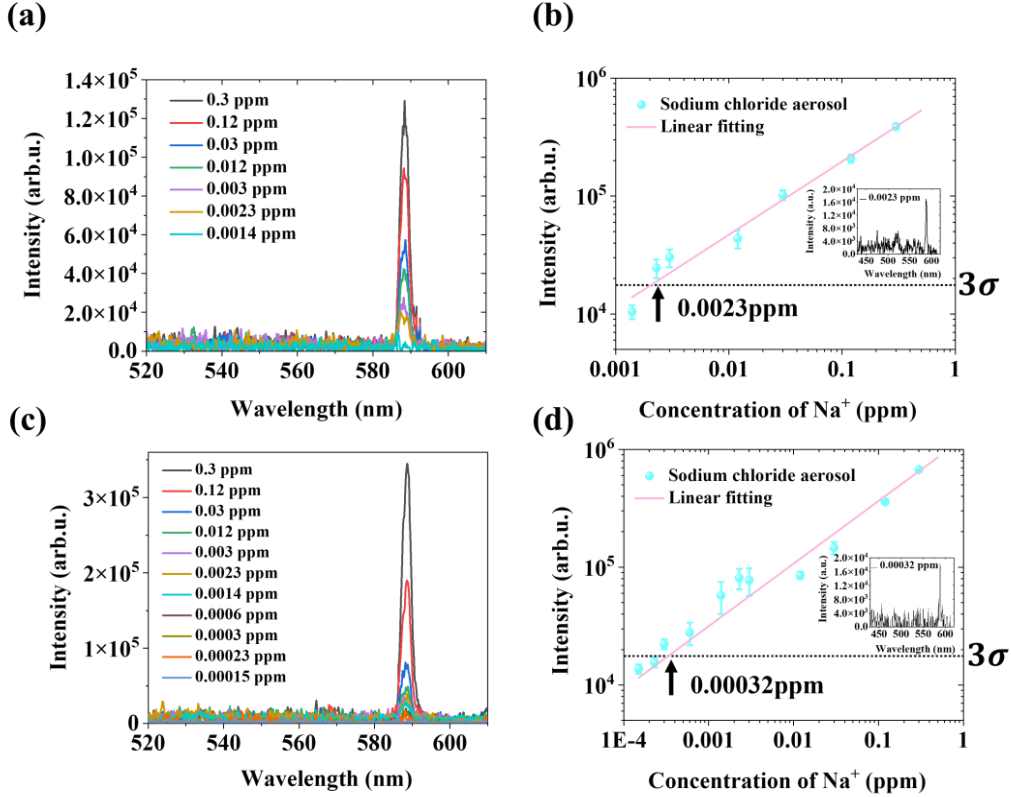

**Fig. 8.** (a) Fluorescence spectra of NaCl aerosol with different concentrations when the laser beam symmetrically strikes on the concave mirror (see Figure 2a); (b) integral intensity of the fingerprint fluorescence peak of NaCl aerosol as a function of Na+ concentrations which is summarized using the data in (a); (c) fluorescence spectra of NaCl aerosol with different concentrations when the laser beam non-symmetrically strikes on the concave mirror (see Figure 3a);; (d) integral intensity of the fingerprint fluorescence peak of NaCl aerosol as a function of Na+ concentrations which is summarized using the data in (c).

Due to the limited size of the laboratory, we can only achieve the experimental verification at a distance of 30 m. In practical ambient air environment, air turbulence may have many adverse effects on laser transmission when the detection distance reaches the order of kilometers, such as the beam cross-sectional intensity distortion, pulse temporal profile distortion and phase fluctuation[34, 35]. Fortunately, many spatiotemporal modulation methods have been developed to overcome the turbulence induced beam distortion[36, 37]. Furthermore, our previous experimental results also show that the air turbulence may even improve the limit of detection of aerosol due to the generation of multiple optical filaments, i.e. the increase of the filament number[38].

## 4. Conclusion

In this work, a novel method which can eliminate the astigmatism of the large-aperture off-axis femtosecond laser filamentation system is proposed. Astigmatism in off-axis system is mainly caused by the non-symmetric wavefront distortion when the beam is obliquely incident into the optical setup. By introducing additional non-symmetricity in the off-axis optical setup, the astigmatism is almost completely eliminated. As a result, the limit of detection of aerosol by femtosecond laser filament induced fluorescence is significantly improved, which achieves a record of sub-ppb aerosol detection limit in a distance of 30 m by millijoule laser pulse. The results in this paper solve one of the key problems hindering the femtosecond laser filament remote sensing, which greatly promotes the development of the related research fields.

**Acknowledgments**

This work was supported by National Key Research and Development Program of China (2018YFB0504400).; Fundamental Research Funds for the Central Universities (63223052).

**Compliance with ethics guidelines**

Jiewei Guo, Zhi Zhang, Nan Zhang, Binpeng Shang, Jiayun Xue, Yuezheng Wang, Shishi Tao, Bofu Xie, Lanjun Guo,Lie Lin and Weiwei Liu declare that they have no conflict of interest or financial conflicts to disclose.